\documentclass[12pt]{iopart}

\usepackage[utf8]{inputenc}
\usepackage[T1]{fontenc}
\usepackage{color}
\usepackage{epsfig,amsfonts}
\usepackage{graphicx}
\usepackage{float}
\usepackage{epstopdf}
\usepackage{blindtext}
\usepackage{bbold}

\def\be{\begin{equation}}
\def\ee{\end{equation}}

\def\bea{\begin{eqnarray}}
\def\eea{\end{eqnarray}}

\def\ben{\begin{enumerate}}
\def\een{\end{enumerate}}

\newcommand{\ket}[1]{\left| #1 \right>}

\def\bea{\begin{eqnarray}}
\def\eea{\end{eqnarray}}

\def\ket#1{\mathinner{|{#1}\rangle}}

\begin{document}

\title[Quadratic relations in Richardson-Gaudin models]{Quadratic operator relations and Bethe equations for spin-1/2 Richardson-Gaudin models}
\author{Claude Dimo, Alexandre Faribault}
\address{ Universit\'e de Lorraine, CNRS, LPCT, F-54000 Nancy, France}
\ead{alexandre.faribault@univ-lorraine.fr}
\begin{abstract}
In this work we demonstrate how one can, in a generic approach, derive a set of $N$ simple quadratic Bethe equations for integrable Richardson-Gaudin (RG) models built out of $N$ spins-1/2. These equations depend only on the $N$ eigenvalues of the various conserved charges so that any solution of these equations defines, indirectly through the corresponding set of eigenvalues, one particular eigenstate.

The proposed construction covers the full class of integrable RG models of the XYZ (including the subclasses of XXZ and XXX models) type realised in terms of spins-1/2, coupled with one another through $\sigma_i^x \sigma_j^x $, \  $\sigma_i^y \sigma_j^y $, \ $\sigma_i^z \sigma_j^z $ terms, including, as well, magnetic field-like terms linear in the Pauli matrices.

The approach exclusively requires integrability, defined here only by the requirement that $N$ conserved charges $R_i$ (with $i = 1,2 \dots N$)  such that $\left[R_i,R_j\right] =0 \   (\forall \ i,j)$ exist . The result is therefore valid, and equally simple, for models with or without $U(1)$ symmetry, with or without a properly defined pseudo-vacuum as well as for models with non-skew symmetric couplings.

\end{abstract}

\date{\today}

\section{Introduction}

It has been observed, at least since the work of Babelon and Talalaev on the quantum Jaynes-Cummings-Gaudin model \cite{babelon}, that certain integrable Gaudin models allow for the construction of quadratic Bethe equations, depending on the eigenvalues of the various conserved charges instead of the traditional Bethe roots. This fact has since been vastly exploited in numerical work \cite{faribaultgritsev1,faribaultgritsev2,straeter,faribaultschuricht1,faribaultschuricht2,claeysfloquet,claeysvariational,claeystopo} due to the simplicity these quadratic equations offer when numerically looking for their solutions. In fact,  the traditional Bethe equations defined in terms of a set of Bethe roots are, in RG models, plagued with cancelling divergences. The eigenvalue-based approach avoids completely these complications, making their direct numerical solutions a much simpler numerical task. The recent emergence of determinant representations, expressed directly in terms of the eigenvalues, for the norms, scalar products and form factors of local operators \cite{faribaultschurichtdet,claeysdet,claeysdet2,claeysdet3,tschirhartdet1,tschirhartdet2,tschirhartdet3} has also been a major step forward in the numerical use of integrability for the study of the static and dynamical physical properties of these systems.

It was recently pointed out that the existence of quadratic eigenvalue-based equations for models built out of spins-1/2, translate into quadratic relations between the conserved operators themselves  \cite{claeystopo,linksscipost,linksproc}. A generalisation to cubic relations between conserved operators has also appeared for spin 1 models \cite{linkstalk}. While these relations have, so far, been constructed on a case by case basis, this work aims at providing a generic framework for the existence of such quadratic relations in spins-1/2 RG-models.

Working on the operator level, we show that, in this general class of models, integrability itself (having the $N$ operators which all commute with one another) implies that these commuting operators are necessarily linked through a closed set of quadratic operator relations. This then allows one to explicitly construct a set of $N$ quadratic Bethe equations obeyed by the conserved charges' eigenvalues whose solution gives access to spectrum of the problem without any necessity to build an explicit Bethe Ansatz solution.

The fact that the approach presented in this work defines Bethe equations without making reference to its Bethe Ansatz solvability (i.e. without any need to build the eigenstates) presents a clear advantage for models without $U(1)$ symmetry. One such example are the XYZ models, related to 8-vertex models of statistical physics, for which the $U(1)$ symmetry  associated to rotational invariance in the $x-y$ plane found in XXZ or XXX models (6-vertex), is longer present. In such a case, a proper highest-weight state which serves as the pseudo-vacuum (reference state)  for the Algebraic Bethe Ansatz is no longer simply defined \cite{sklyaninXYZ}. Moreover, cases for which the $U(1)$ symmetry is explicitly broken by an in-plane component of the magnetic field can also make such a proper pseudo-vacuum non-existant \cite{claeystopo,tschirhartdet3}. A variety of techniques have been developed over the years to deal with models lacking such an explicitly known reference state: Separation of Variables \cite{sklyaninsov,sklyaninelliptic,hikami,sklyanintrends,Niccoliff,Niccoliap},  Off-Diagonal Bethe Ansatz \cite{ hao,wangbook,ODBA2,ODBA3,ODBA4} or the Modified Bethe Ansatz \cite{claeystopo,tschirhartdet3,crampe,belliardnou1,belliardcrampe,belliard,belliard1,belliard2,belliard3}. However, by circumventing the construction of eigenstates, the approach proposed in this work makes all the possible cases equally simple, at least when exclusively looking for the eigenenergies.

This article begins with section \ref{integrability} where we define the class of integrable models treated in this work and derive the set of conditions on the coupling constants which are imposed by integrability. Section \ref{quadraticity} then defines the ensemble of requirements for quadratic relations to exist between the various conserved operators. In section \ref{equivalence}, the equivalence between the two sets of conditions is established, demonstrating that integrability itself is a sufficient condition for such quadratic relations to exist and that, simply defining the set of conserved charges allows one to explicitly construct the quadratic Bethe equations. Finally, in section \ref{knowncases}, known cases from the literature on spins-1/2 systems are verified again and shown to be particular cases of the general result found in this work.

\section{The models and their integrability}
\label{integrability}

We focus in this work, on spin-1/2 models defined by the following commuting conserved charges, quadratic in Pauli matrices:

\bea
R_i = \vec{B}_i \cdot \vec{\sigma}_i + \sum_{k \ne i}^N \sum_{\alpha=x,y,z} \Gamma^\alpha_{i,k} \  \sigma^\alpha_i\sigma^\alpha_k. 
\label{consdef}
\eea

While this readily excludes certain more generic integrable models containing Dzyaloshinskii-Moriya-like terms coupling $\sigma_i^\alpha$ and $\sigma_j^\beta$, it covers the complete class of "traditional" XYZ-RG-models.

One should note that, in this work, greek superscripts will systematically be used to denote orientations (taking three possible values: $x,y,z$) while latin subscripts will be used to label individual spins therefore taking the values $1,2,3 \dots N$. Requiring that the conserved charges (\ref{consdef}) commute for all $i,j$ does impose a set of constraints, which we call integrability constraints since they simply enforce the required commutation rules needed to define an integrable model of this type. The approach used here to find the integrability constraints has appeared in many works concerning Gaudin models \cite{amico,dukelsky} starting with Gaudin himself \cite{gaudin76,gaudinbook,gaudinbook2}. Straightforwardly, the constraints are found by explicitly enforcing that the commutators be equal to zero:
\bea
\left[R_i,R_j \right]= 0.
\eea

These commutators are simple to compute using the commutation rules of Pauli matrices
\bea
\left[\sigma^\alpha_i,\sigma^\beta_j \right] = 2 i \ \delta_{i j}  \ \epsilon_{\alpha \beta \gamma} \ \sigma^\gamma_i,
\eea

\noindent where $\epsilon$ is the Levi-Civita antisymmetric symbol.

Since the terms, linear in Pauli matrices: $\vec{B}_i \cdot \vec{\sigma}_i $ and $\vec{B}_j \cdot \vec{\sigma}_j $, commute because they involve distincts spins, the commutator can be expanded as:

\bea
\left[R_i,R_{j (\ne i)}\right] 
&=&
 \sum_{\beta,\gamma} B^\gamma_i \Gamma^{\beta}_{j i}  \left[\sigma_i^\gamma , \sigma_i^\beta
\right]\sigma_j^\beta+\sum_{\alpha,\gamma}B^\gamma_j \Gamma^{ \alpha}_{i j}  \sigma_i^\alpha \left[ \sigma_j^\alpha,  \sigma_j^\gamma 
\right]
\nonumber\\ &&+  \sum_{k\ne i,j }^N \sum_{\alpha,\beta}\Gamma^{\alpha}_{i k}  \Gamma^{\beta}_{j k}\sigma_i^\alpha \sigma_j^\beta \left[\sigma_k^\alpha,\sigma_{k}^\beta
\right]
+\sum_{k\ne i,j }^N\sum_{\beta,\gamma}\Gamma^{\gamma }_{i k} \Gamma^{\beta}_{j i}   \left[ \sigma_i^\gamma,  \sigma_{i}^\beta
\right]\sigma_j^\beta \sigma_k^\gamma 
\nonumber\\ &&+  \sum_{k\ne j,i }^N \sum_{\alpha,\gamma}\Gamma^{\alpha}_{i j} \Gamma^{\gamma}_{j k} \sigma_i^\alpha \left[\sigma_j^\alpha,  \sigma_j^\gamma
\right]
 \sigma_{k}^\gamma,
 \label{intcond}
\eea

\noindent using the fact that $\left[\sigma_i^\alpha\sigma_j^\alpha, \sigma_j^\beta\sigma_i^\beta\right] =0$. One finds  quadratic and cubic terms in Pauli matrices and each of the coefficients in front of them needs to explicitly cancel out for the model to be integrable. This leads to a series of algebraic relations  between the couplings $\Gamma$ and the various "magnetic fields" $B$. From the quadratic terms one finds, for any permutation of the sets $\{\alpha,\beta,\gamma\} = \{x,y,z\}$, that:
\bea
 B^\gamma_i  \Gamma^{\beta}_{j i} +  B^\gamma_j  \Gamma^{\alpha}_{i j}  = 0 \ \ \ \forall \ \alpha \ne \beta \ne \gamma,
 \label{intbgamma}
\eea
\noindent while the cubic terms lead, for each permutation, to a Gaudin equation imposed on the couplings:
\bea
\Gamma^{\alpha}_{i k}  \Gamma^{\beta}_{j k} -\Gamma^{\gamma }_{i k} \Gamma^{\beta}_{j i}-\Gamma^{\alpha}_{i j} \Gamma^{\gamma}_{j k} = 0    \ \ \ \forall \ \alpha \ne \beta \ne \gamma.
 \label{intgammagamma}
\eea

These two restrictions, defining integrability, are the only requirements which we will impose on the models in that no $U(1)$-symmetry nor existence of an adequate pseudo-vaccum will be demanded.  The antisymmetry of the couplings $\Gamma^\alpha_{ij} = - \Gamma^\alpha_{ji} $ is {\bf NOT} going to be imposed here either, as is frequently the case following Gaudin's three proposed antisymmetric solutions: rational (XXX), trigonometric (XXZ) and elliptic (XYZ) \cite{gaudin76} also defining the Belavin-Drinfel'd classification of solutions \cite{belavin}. While Gaudin mentioned explicitly that antisymmetry needs not to be imposed \cite{gaudin76}, models for which it is not have been mostly studied in more recent years. For example, Balantekin et al. \cite{balantekin} has defined such non-skew-symmetric integrable models for which $\Gamma^\alpha_{ij} = -\Gamma^\alpha_{ji} - 2 q$ with $q$ any real constant. A large body of work by Skrypnik \cite{skrypnyk1,skrypnyk2,skrypnyk3,skrypnyk4,skrypnyk5,skrypnyk6,skrypnyk7,skrypnyk8} as well as a recent paper by Links \cite{linksnonskew} have have also dealt with models which are not within the usual antisymmetric classes of solutions to the Gaudin equation. In this work, antisymmetry is never imposed so all of the possible non-skew symmetric integrable models defined by conserved charges of the form (\ref{consdef}) are also naturally included in the approach.

\section{Quadratic operator relations}
\label{quadraticity}

In a few specific cases \cite{claeystopo,tschirhartdet3,linksscipost, linksproc,linkstalk}, it was shown explicitly that the eigenvalues $r_i$ of the conserved charges $(R_i)$ can be related by quadratic equations linking $r_i^2$ to a linear combination of all the $r_k$. Considering that these relations hold for the eigenvalues associated to each of the eigenstates and that, in the common eigenbasis, the conserved charges $R_i$ are all diagonal operators, the quadratic relations between the eigenvalues also hold for the operators themselves. 

We therefore aim to see whether some (or all) of the integrable models studied here can obey, on the operator level, the following relation:

\bea
R_i^2 = \sum_{j\ne i} C_{ij} R_j + K_i
\eea

\noindent with $K_i$ and $C_{ij}$ a set of constants to be determined for a given model.

For such a relation to be valid, the conserved charges again need to obey a set of constraints which one can find by first squaring the conserved charge $R_i$. The simplest way to do so is to compute the anticommutator of $R_i$ with itself:

\bea
R_i^2 = \frac{1}{2} \left[R_i,R_i\right]_+
\eea

\noindent since the anticommutator of Pauli matrices is known to be given by:

\bea
\frac{1}{2} \left[\sigma^\alpha_i,\sigma^\beta_i\right]_+ = \frac{1}{2}\sigma^\alpha_i \sigma^\beta_i+ \frac{1}{2}\sigma^\beta_i\sigma^\alpha_i= \delta_{\alpha \beta} \ \mathbb{1}.
\eea

In terms of the linear and quadratic (in Pauli matrices) terms in the conserved charge $R_i$:
\bea
R^l_i &=& \sum_{\alpha} B^\alpha_i \sigma_i^\alpha 
\ \ \ \ \ \ \
R^q_i =  \sum_{\alpha} \sum_{k\ne i }^N \Gamma^{\alpha}_{i k} \sigma_i^\alpha\sigma_k^\alpha
\eea

 \noindent one has:
\bea
R_i^2 = \frac{1}{2} \left[R^l_i , R^l_i \right]_+ +\left[R^l_i , R^q_i \right]_+ + \frac{1}{2} \left[R^q_i , R^q_i \right]_+.
\label{anticosquare}
\eea

The first anticommutator in eq. (\ref{anticosquare}), is easily shown to simply give a constant:

\bea
 \frac{1}{2} \left[R^l_i , R^l_i \right]_+ =   \left(\sum_{\alpha} (B^\alpha_i)^2 \right) \mathbb{1},
\eea

\noindent while the second one is straightforwardly shown to be given by:

\bea
\left[R^l_i , R^q_i \right]_+ &=&   \sum_{k\ne i }^N  \sum_{\alpha} 2 B^\alpha_i \Gamma^{\alpha}_{i k}  \sigma_k^\alpha,
\eea

\noindent contributing, to $R_i^2$, linear terms in Pauli matrices involving exclusively the spins of index $k \ne i$. The remaining anticommutator can be expanded as
\bea
&& \frac{1}{2} \left[R^q_i , R^q_i \right]_+ = 
 \sum_{\alpha,\gamma} \sum_{k\ne i }^N\sum_{k'\ne i }^N \Gamma^{\alpha}_{i k}\Gamma^{\gamma }_{i k'} \left[ \sigma_i^\alpha\sigma_k^\alpha, \sigma_i^\gamma\sigma_{k'}^\gamma\right]_+
\nonumber\\
&=&\sum_{\alpha,\gamma} \sum_{k\ne i }^N\sum_{k'\ne i,k }^N   \frac{\Gamma^{\alpha}_{i k}\Gamma^{\gamma }_{i k'}}{2}\left[ \sigma_i^\alpha\sigma_k^\alpha, \sigma_i^\gamma\sigma_{k'}^\gamma\right]_+
 +\sum_{\alpha,\gamma} \sum_{k\ne i }^N \frac{ \Gamma^{\alpha }_{i k} \Gamma^{\gamma }_{i k}}{2}\left[ \sigma_i^\alpha\sigma_k^\alpha, \sigma_i^\gamma\sigma_{k}^\gamma\right]_+
 \nonumber\\
\eea

\noindent  by splitting it into its $k \ne k'$ and $k = k'$ terms. The first term has $k'\ne k \ne i$ and therefore, since Pauli matrices associated with distinct spins all commute with one another, gives:

\bea
\sum_{\alpha,\gamma} \sum_{k\ne i }^N\sum_{k'\ne i,k }^N   \frac{\Gamma^{\alpha }_{i k}\Gamma^{\gamma}_{i k'}}{2}\left[ \sigma_i^\alpha\sigma_k^\alpha, \sigma_i^\gamma\sigma_{k'}^\gamma\right]_+
&=&
\sum_{\alpha,\gamma} \sum_{k\ne i }^N\sum_{k'\ne i,k }^N   \Gamma^{\alpha}_{i k}\Gamma^{\gamma}_{i k'} \delta_{\alpha,\gamma} \sigma_k^\alpha\sigma_{k'}^\gamma
\nonumber\\
&=&
\sum_{\alpha} \sum_{k\ne i }^N\sum_{k'\ne i,k }^N   \Gamma^{\alpha}_{i k}\Gamma^{\alpha}_{i k'}  \sigma_k^\alpha\sigma_{k'}^\alpha
\nonumber\\
\eea

The remaining term $k'=k$ can be, using the known commutators, anticommutators and product:
\bea
\sigma^\alpha_j\sigma^\beta_j  &=& \delta_{\alpha \beta}\mathbb{1} + i \epsilon_{\alpha \beta \gamma} \sigma^\gamma_j,
\eea 

\noindent rewritten as:
\bea
 \left[ \sigma_i^\alpha\sigma_k^\alpha, \sigma_i^\gamma\sigma_{k}^\gamma\right]_+
&=& \sigma_i^\alpha\sigma_k^\alpha \sigma_i^\gamma\sigma_{k}^\gamma + \sigma_i^\gamma\sigma_{k}^\gamma\sigma_i^\alpha\sigma_k^\alpha
\nonumber\\
&=&
\sigma_i^\alpha\sigma_i^\gamma \sigma_k^\alpha \sigma_{k}^\gamma + \sigma_i^\gamma\sigma_i^\alpha\sigma_k^\alpha \sigma_{k}^\gamma +
\sigma_i^\gamma\sigma_i^\alpha\left[\sigma_{k}^\gamma,\sigma_k^\alpha\right] 
\nonumber\\
&=&
\left[\sigma_i^\alpha,\sigma_i^\gamma\right]_+ \sigma_k^\alpha \sigma_{k}^\gamma + 
\sigma_i^\gamma\sigma_i^\alpha\left[\sigma_{k}^\gamma,\sigma_k^\alpha\right]
\nonumber\\
&=&
2 \delta_{\alpha \gamma} \sigma_k^\alpha \sigma_{k}^\gamma + 
2 i \epsilon_{\gamma,\alpha,\beta} \sigma_i^\gamma\sigma_i^\alpha
\sigma_{k}^\beta
\nonumber\\
&=&
2 \delta_{\alpha\gamma} 
\left(\delta_{\alpha \gamma} \mathbb{1} + i \epsilon_{\alpha,\gamma,\beta}  \sigma_k^\beta\right)
 + 
2 i \epsilon_{\gamma,\alpha,\beta}
\left(\delta_{\gamma\alpha} \mathbb{1} + i \epsilon_{\gamma,\alpha,\beta}  \sigma_i^\beta\right)
\sigma_{k}^\beta
\nonumber\\
&=&
2 \delta_{\alpha,\gamma} 
 \mathbb{1} 
 - 
2  
  \sigma_i^\beta
\sigma_{k}^\beta (\epsilon_{\alpha \gamma \beta})^2,
\eea 

\noindent  leading to a constant term for $\alpha = \gamma$ and a quadratic term which couples spin $i$ and spin $k$ along the $\beta \ne (\alpha,\gamma)$ direction. Therefore, the remaining term is given by:

\bea
\sum_{\alpha,\gamma} \sum_{k\ne i }^N \frac{ \Gamma^{\alpha }_{i k} \Gamma^{\gamma }_{i k}}{2}\left[ \sigma_i^\alpha\sigma_k^\alpha, \sigma_i^\gamma\sigma_{k}^\gamma\right]_+
&=&
\sum_{\alpha,\gamma} \sum_{k\ne i }^N  \Gamma^{\alpha }_{i k} \Gamma^{\gamma }_{i k}
\left(
 \delta_{\alpha,\gamma} 
 \mathbb{1} 
 -   
  \sigma_i^\beta
\sigma_{k}^\beta (\epsilon_{\alpha \gamma \beta})^2
\right).
\eea

In the end, one finds:
\bea
R_i^2 &=&  \left(\sum_{\alpha} (B^\alpha_i)^2 + \sum_{\alpha} \sum_{k\ne i }^N \left( \Gamma^{\alpha}_{i k} \right)^2\right) \mathbb{1}
+  \sum_{\alpha}\sum_{k\ne i }^N 2 B^\alpha_i \Gamma^{\alpha}_{i k} \sigma_k^\alpha
\nonumber\\ &+&
\sum_{\alpha} \sum_{k\ne i }^N\sum_{k'\ne i,k }^N   \Gamma^{ \alpha}_{i k}\Gamma^{\alpha}_{i k'} \sigma_k^\alpha\sigma_{k'}^\alpha
- 2 \sum_{\alpha } \sum_{k\ne i }^N  \left( \Gamma^{\beta}_{i k} \Gamma^{\gamma }_{i k} \right)
  \sigma_i^\alpha
\sigma_{k}^\alpha,
\eea

\noindent where, in the last term, $\beta$ and $\gamma$ are the two directions perpendicular to $\alpha$.

There remains to be seen under which conditions the resulting squared conserved charge can be rewritten as a linear combination of the other conserved charges. Such a generic linear combination can be written as:

\bea
&&
\sum_{k \ne i} C_{ik} R_k + K_i \cdot \mathbb{1} = \sum_{\alpha}\sum_{k \ne i} C_{ik}
B^\alpha_k \sigma_k^\alpha +  \sum_{\alpha}\sum_{k \ne i}   \sum_{k'\ne k}^N C_{ik} \Gamma^{\alpha}_{k k' } \sigma_k^\alpha\sigma_{k'}^\alpha+ K_i \cdot \mathbb{1}
\nonumber\\
&& = \sum_{\alpha}\sum_{k \ne i}  C_{ik}
B^\alpha_k \sigma_k^\alpha + \sum_{\alpha}  \sum_{k \ne i}   \sum_{k'\ne i,k}^N C_{ik} \Gamma^{\alpha}_{k k' } \sigma_k^\alpha\sigma_{k'}^\alpha
+ \sum_{\alpha} \sum_{k \ne i}   C_{ik} \Gamma^{\alpha}_{k i } \sigma_{i}^\alpha\sigma_k^\alpha+ K_i \cdot \mathbb{1}
\nonumber\\
\eea

Demonstrating that such a linear combination exists amounts to showing that constants $C_{ij}$ can be defined in a way which is consistent, term by term, with the previous expression for $R_i^2$. Namely, one needs:

\bea
K_i =  \left(\sum_{\alpha} (B^\alpha_i)^2 + \sum_{\alpha} \sum_{k\ne i }^N \left( \Gamma^{\alpha}_{i k} \right)^2\right) 
\label{constants}
\\
C_{ik} B^\alpha_k  = 2 B^\alpha_i \Gamma^{\alpha }_{i k}
\label{bgamma} 
\\
 C_{ik} \Gamma^{\alpha}_{k i }  =  -2  \Gamma^{\beta}_{i k} \Gamma^{\gamma }_{i k}
\label{gammagammai}
\\
C_{ik} \Gamma^{\alpha }_{k k' } + C_{ik'} \Gamma^{ \alpha}_{k' k } =  2\Gamma^{ \alpha}_{i k}\Gamma^{\alpha}_{i k'} \  \ \ \forall  \ k' > k \ \ \ (k,k' \ne i),
\label{gammagammak}
\eea

\noindent where the last equation covers every pair of {\bf distinct} $k,k'$ indices.

\section{Integrability and quadraticity}
\label{equivalence}

The first integrability condition (\ref{intbgamma}) found previously:

\bea
 B^\gamma_i  \Gamma^{\beta}_{k i} +  B^\gamma_k  \Gamma^{\alpha}_{ik}  = 0 \to 
 B^\gamma_i  \Gamma^{\beta}_{k i}   = -    B^\gamma_k  \Gamma^{\alpha}_{ik}
 \ \ \ \forall \ \alpha \ne \beta \ne \gamma
 \label{intcond1}
\eea

\noindent can be used to prove that, if it is respected, integrability guarantees that the three equations (\ref{bgamma}) (along $\alpha = x,y,z$ ): 

\bea
C_{ik} B^\alpha_k  = 2 B^\alpha_i \Gamma^{\alpha }_{i k}
\eea

\noindent are all consistent, i.e all lead to the same $C_{ik}$. This of constistency equations:

\bea
\frac{ B^\alpha_i}{B^\alpha_k } \Gamma^{\alpha }_{i k} = \frac{ B^\beta_i}{B^\beta_k } \Gamma^{\beta }_{i k} \ \to \ 
B^\alpha_iB^\beta_k  \Gamma^{\alpha }_{i k} = B^\beta_i B^\alpha_k  \Gamma^{\beta }_{i k},
\eea

\noindent  are indeed respected if the integrability condition is met. Indeed, if condition (\ref{intcond1}) is verified, the previous consistency equation can be rewritten as:

\bea
B^\alpha_i \left(- B^\beta_i \Gamma^{\gamma }_{ ki} \right)
 = B^\beta_i \left(- B^\alpha_i \Gamma^{\gamma }_{ ki}\right),
\eea

\noindent proving it is systematically true.

On the other hand, we further need to prove that (\ref{gammagammai}) and (\ref{gammagammak}) are also consistent with this particular set of constants $C_{ik}   = 2 \frac{B^\alpha_i}{B^\alpha_k} \Gamma^{\alpha }_{i k} $ (which were shown to be equal for any of the three possible directions $\alpha$). The right hand term in eq. (\ref{gammagammak}) can, using the integrability relation (\ref{intcond1}), be written as:

\bea
2\Gamma^{ \alpha}_{i k}\Gamma^{\alpha}_{i k'} = - 2 \frac{B^\gamma_i}{B^\gamma_k}  \Gamma^{\beta}_{k i} \Gamma^{\alpha}_{i k'} =   2 \frac{B^\gamma_i}{B^\gamma_k} \left(-\Gamma^{\beta}_{k k'}  \Gamma^{\gamma}_{i k'} +\Gamma^{\alpha }_{k k'} \Gamma^{\gamma}_{i k}  \right),
\label{partialproof}
\eea

\noindent by using the second integrability condition (\ref{intgammagamma}) in the form:
\bea
\Gamma^{\beta}_{k i} \Gamma^{\alpha}_{i k'} = -\Gamma^{\beta}_{k k'}  \Gamma^{\gamma}_{i k'} +\Gamma^{\alpha }_{k k'} \Gamma^{\gamma}_{i k}.   \eea

The other integrability constraint  (\ref{intcond1}) then allows us to write:

\bea
2\Gamma^{ \alpha}_{i k}\Gamma^{\alpha}_{i k'}  =   2 \frac{B^\gamma_i}{B^\gamma_k} \left(\frac{B^\gamma_k}{B^\gamma_{k'}}\Gamma^{\alpha}_{k' k}  \Gamma^{\gamma}_{i k'} +\Gamma^{\alpha }_{k k'} \Gamma^{\gamma}_{i k}  \right) = 2 \left(\frac{B^\gamma_i}{B^\gamma_{k'}}\Gamma^{\alpha}_{k' k}  \Gamma^{\gamma}_{i k'} + \frac{B^\gamma_i}{B^\gamma_{k}}\Gamma^{\alpha }_{k k'} \Gamma^{\gamma}_{i k}  \right).\nonumber\\
\eea  

For the constants $C_{ij}$ as they have been defined in (\ref{bgamma}), this last equality becomes:

\bea
2\Gamma^{ \alpha}_{i k}\Gamma^{\alpha}_{i k'} &=& C_{ik'} 
\Gamma^{\alpha}_{k k'}  
+ C_{ik}\Gamma^{\alpha}_{k k'}
\eea 

\noindent proving that the integrability conditions are sufficient to insure that the quadraticity condition (\ref{gammagammak}) is met. It now simply remains to verify that the last condition (\ref{gammagammai}) is also verified. For the constants $C_{ij}$ defined by (\ref{bgamma}), one finds:

\bea
C_{ik} \Gamma^{\alpha}_{k i }  = 2 \frac{B^\gamma_i}{B^\gamma_k} \Gamma^{\gamma }_{i k} \Gamma^{\alpha}_{k i }=  2 \frac{B^\gamma_i}{B^\gamma_k}\Gamma^{\alpha}_{k i }   \Gamma^{\gamma }_{i k}  =  -2  \Gamma^{\beta}_{i k} \Gamma^{\gamma }_{i k},
\eea

\noindent where the integrability condition (\ref{intcond1}) has been used. This last equality confirms that integrability itself is also sufficient for the last remaining quadraticity condition (\ref{gammagammai}) to be respected. 

It has therefore been proven that, for any integrable RG model defined by conserved charges (\ref{consdef}) which all commute with one another, the square of the conserved charges can be written as the following linear combination of the other conserevd charges:

\bea
R_i^2 = - 2 \sum_{j \ne i} \frac{\Gamma^{\alpha}_{ij}\Gamma^{\gamma}_{ij}}{\Gamma^\beta_{ji}} R_j +  \sum_{\alpha} \left(B^\alpha_i\right)^2 +\sum_\alpha \sum_{k\ne i} \left(\Gamma^\alpha_{ik}\right)^2.
\label{quadop}
\eea

Here we chose to use (\ref{gammagammai}) to write the constants $C_{ij} = -2\frac{\Gamma^{\alpha}_{ij}\Gamma^{\gamma}_{ij}}{\Gamma^\beta_{ji}} $, but they can also be equivalently written as $C_{ij} = 2 \frac{B_i^\alpha \Gamma^\alpha_{ik}}{B_k^\alpha}$. Since the quadratic relation is valid on the operator level, it is also trivially valid for the set of eigenvalues $r_i$ associated with one of the eigenstates. This therefore finally provides the following set of quadratic Bethe equations:

\bea
r_i^2 = - 2 \sum_{j \ne i}^N \frac{\Gamma^{\alpha}_{ij}\Gamma^{\gamma}_{ij}}{\Gamma^\beta_{ji}} \ r_j +  \sum_{\alpha} \left(B^\alpha_i\right)^2 +\sum_\alpha  \sum_{k\ne i} \left(\Gamma^\alpha_{ik}\right)^2,
\eea

\noindent whose set of solutions will define the complete energy eigenspectrum of the model.

\section{Known cases}
\label{knowncases}

While the main result obtained in this work applies to a much broader set of models, we show explicitly in this section how it allows one to reproduce the known specific cases for which  quadratic operator relations have been published previously.

\subsection{XXX-Richardson-Gaudin} 

Such a quadratic operator relation was explicitly proven by Links in \cite{linksscipost} using a permutation operator representation of the XXX-RG-models. Defining conserved charges as:

\bea
T_i = B \sigma^z_i + \sum_{j\ne i}\frac{\mathcal{P}_{ij}-1}{\epsilon_i-\epsilon_j} = \alpha \sigma^z_i + \frac{1}{2} \sum_{j\ne i}\frac{\vec{\sigma}_{i}\cdot \vec{\sigma}_{j}-1}{\epsilon_i-\epsilon_j}  
\eea

\noindent which were shown to obey the quadratic relations:

\bea
T_i^2=B^2 - \sum_{j \ne i} \frac{T_i - T_j}{\epsilon_i - \epsilon_j}.
\label{tquad}
\eea

This specific example corresponds, in our general formula  (\ref{quadop}), to the case defined by isotropic and skew-symmetric couplings given by:

\bea
\Gamma_{ij}^\alpha = \frac{1}{2}\frac{1}{\epsilon_i-\epsilon_j}  \ \ \ \forall \ \ \ i \ne j \ \ \ \forall \ \ \alpha = x,y,z.
\eea
\noindent and magnetic field terms given by
\bea
B^\alpha_i = B \delta_{\alpha z},
\eea

\noindent for which the general equation (\ref{quadop}) can be rewritten, using the fact that $\Gamma^\beta_{ji} = - \Gamma^\beta_{ij}$, as:

\bea
R_i^2 =   \sum_{j \ne i} \frac{R_j }{\epsilon_i-\epsilon_j} +  B^2 + \frac{3}{4} \sum_{j\ne i}\frac{1}{(\epsilon_i-\epsilon_j)^2}
\eea

Since $\displaystyle  R_i = T_i + \frac{1}{2}\sum_{j \ne i} \frac{1}{\epsilon_i - \epsilon_j}$ it does finally become equivalent to eq. (\ref{tquad}):

\bea
 &&T_i^2 + \sum_{j \ne i} \frac{T_i}{\epsilon_i - \epsilon_j}
  + \frac{1}{4}\sum_{j \ne i} \sum_{k \ne i} \frac{1}{\epsilon_i - \epsilon_j}\frac{1}{\epsilon_i - \epsilon_k}
 \nonumber\\ &&= B^2 +
 \sum_{j\ne i} \frac{T_j}{\epsilon_i-\epsilon_j} + \frac{1}{2} \sum_{j\ne i} \sum_{k \ne j} \frac{1}{\epsilon_i-\epsilon_j} \frac{1}{\epsilon_j - \epsilon_k}
 +\frac{3}{4} \sum_{j\ne i}\frac{1}{(\epsilon_i-\epsilon_j)^2}
 \nonumber\\
 && T_i^2 = B^2 -
 \sum_{j\ne i} \frac{T_i - T_j}{\epsilon_i-\epsilon_j},
\eea

\noindent since

\bea
 - \frac{1}{4}\sum_{j \ne i} \sum_{k \ne i} \frac{1}{\epsilon_i - \epsilon_j}\frac{1}{\epsilon_i - \epsilon_k}
+ \frac{1}{2} \sum_{j\ne i} \sum_{k \ne j} \frac{1}{\epsilon_i-\epsilon_j} \frac{1}{\epsilon_j - \epsilon_k}
 +\frac{3}{4} \sum_{j\ne i}\frac{1}{(\epsilon_i-\epsilon_j)^2} =0.\nonumber\\
\eea

This last statement can be easily proven:

\bea
&& - \frac{1}{4}\sum_{j \ne i} \sum_{k \ne i} \frac{1}{\epsilon_i - \epsilon_j}\frac{1}{\epsilon_i - \epsilon_k}
+ \frac{1}{2} \sum_{j\ne i} \sum_{k \ne j} \frac{1}{\epsilon_i-\epsilon_j} \frac{1}{\epsilon_j - \epsilon_k}
 +\frac{3}{4} \sum_{j\ne i}\frac{1}{(\epsilon_i-\epsilon_j)^2} 
 \nonumber\\ && =
 - \frac{1}{4}\sum_{j \ne i} \sum_{k \ne i,j} \frac{1}{\epsilon_i - \epsilon_j}\frac{1}{\epsilon_i - \epsilon_k}
+ \frac{1}{2} \sum_{j\ne i} \sum_{k \ne i,j} \frac{1}{\epsilon_i-\epsilon_j} \frac{1}{\epsilon_j - \epsilon_k}
 \nonumber\\ && =
  \frac{1}{4}\sum_{j \ne i} \sum_{k \ne i,j} \left[\frac{2}{\epsilon_i - \epsilon_j}\frac{1}{\epsilon_j - \epsilon_k}-\frac{1}{\epsilon_i - \epsilon_j}\frac{1}{\epsilon_i - \epsilon_k}\right]
  \nonumber\\ && =
  \frac{1}{4}\sum_{j \ne i} \sum_{k \ne i,j} \left[\frac{2}{\epsilon_i - \epsilon_j}\frac{1}{\epsilon_j - \epsilon_k}-\left(
  \frac{1}{\epsilon_i - \epsilon_j}-\frac{1}{\epsilon_i - \epsilon_k}\right)\frac{1}{\epsilon_j-\epsilon_k}\right]
\nonumber\\ && =
  \frac{1}{4}\sum_{j \ne i} \sum_{k \ne i,j} \left[\frac{1}{\epsilon_i - \epsilon_j}\frac{1}{\epsilon_j - \epsilon_k}+\frac{1}{\epsilon_i - \epsilon_k}\frac{1}{\epsilon_j-\epsilon_k}\right] = 0,
\eea

\noindent since the second term, under the exchange of the summation indices $k,j$, is indeed equal to the first one up to a minus sign.

\subsection{XXZ Richardson-Gaudin}
 
 A second set of similar quadratic equations has also been published in Claeys et al. \cite{claeystopo} in an XXZ case  describing an integrable $p+ip$ superconductor coupled to a particle bath, where conserved charges:

\bea
  \tilde{R}_k 
&=&
\frac{1}{2} \sigma^z_k   + \frac{\gamma}{\epsilon_k} \sigma^x_k -\frac{G}{2}\sum_{{k'} \neq k}^N \left[\frac{\epsilon^2_{k'}}{\epsilon^2_{k} - \epsilon^2_{k'}} \sigma^z_k \sigma^z_{k'} + \frac{\epsilon_{k}\epsilon_{k'}}{\epsilon^2_{k} - \epsilon^2_{k'}} \left(\sigma^x_k \sigma^x_{k'} +\sigma^y_k \sigma^y_{k'}    \right)\right]
\nonumber \\ && + \frac{1}{2} \left(1 + G \sum_{k' \ne k }\frac{\epsilon^2_{k'}}{\epsilon^2_{k} - \epsilon^2_{k'}} \right)
\eea

 \noindent were shown to obey the relations
 
 \bea
  \tilde{R}_k ^2 =  \tilde{R}_k  + \left(\frac{\gamma}{\epsilon_k}\right)^2 + G \sum_{k' \ne k }\epsilon^2_{k'}\frac{  \tilde{R}_k -   \tilde{R}_{k'}}{\epsilon^2_{k} - \epsilon^2_{k'}}.
  \label{quadclaeys}
 \eea

Defining $R_k =   \tilde{R}_k - \displaystyle \frac{1}{2} \left(1 + G \sum_{k' \ne k }\frac{\epsilon^2_{k'}}{\epsilon^2_{k} - \epsilon^2_{k'}} \right)$ and using $\Gamma^x_{kk'} =\Gamma^y_{kk'} = -\frac{G}{2} \frac{\epsilon_{k}\epsilon_{k'}}{\epsilon^2_{k} - \epsilon^2_{k'}} $,$\Gamma^z_{kk'} = -\frac{G}{2} \frac{\epsilon_{k'}^2}{\epsilon^2_{k} - \epsilon^2_{k'}}$ and $B^z_k =\frac{1}{2}, B^x_k =\frac{\gamma}{\epsilon_k}, B^y_k =0 $, the general eq. (\ref{quadop}) involves the constants $-2\Gamma^x_{kk'} \Gamma^y_{kk'}/\Gamma^z_{k'k} = - G \frac{\epsilon_{k'}^2}{\epsilon_{k}^2-\epsilon_{k'}^2}$ and leads to:

\bea
&& \tilde{R}^2_k -  \tilde{R}_k \left(1 + G \sum_{k' \ne k }\frac{\epsilon^2_{k'}}{\epsilon^2_{k} - \epsilon^2_{k'}} \right)  + \frac{1}{4} \left(1 + G \sum_{k' \ne k }\frac{\epsilon^2_{k'}}{\epsilon^2_{k} - \epsilon^2_{k'}} \right) \left(1 + G \sum_{k'' \ne k }\frac{\epsilon^2_{k''}}{\epsilon^2_{k} - \epsilon^2_{k''}} \right)
\nonumber\\
&& = - G \sum_{k' \ne k } 
 \frac{\epsilon_{k'}^2}{\epsilon^2_{k} - \epsilon^2_{k'}} 
 \tilde{R}_{k'} +  \frac{G}{2} \sum_{k' \ne k } 
 \frac{\epsilon_{k'}^2}{\epsilon^2_{k} - \epsilon^2_{k'}} \left(1 + G \sum_{k'' \ne k' }\frac{\epsilon^2_{k''}}{\epsilon^2_{k'} - \epsilon^2_{k''}} \right) + \frac{1}{4} 
\nonumber\\
&& + \left(\frac{\gamma}{\epsilon_k}\right)^2  + \sum_{k' \ne k} \left[2 \left(-\frac{G}{2} \frac{\epsilon_{k}\epsilon_{k'}}{\epsilon^2_{k} - \epsilon^2_{k'}}\right)^2 
 + \left(-\frac{G}{2} \frac{\epsilon_{k'}^2}{\epsilon^2_{k} - \epsilon^2_{k'}}\right)^2
 \right]
 \nonumber\\
 \nonumber\\
 && \tilde{R}^2_k - \tilde{R}_k   -  G \sum_{k' \ne k } \epsilon_{k'}^2
 \frac{\tilde{R}_k - \tilde{R}_{k'} }{\epsilon^2_{k} - \epsilon^2_{k'}} 
 - \left(\frac{\gamma}{\epsilon_k}\right)^2 
\nonumber\\
 && =
-\frac{1}{4} \left(1 + G \sum_{k' \ne k }\frac{\epsilon^2_{k'}}{\epsilon^2_{k} - \epsilon^2_{k'}} \right) \left(1 + G \sum_{k'' \ne k }\frac{\epsilon^2_{k''}}{\epsilon^2_{k} - \epsilon^2_{k''}} \right)+  \frac{1}{4} 
\nonumber\\ && \ +  \frac{G}{2} \sum_{k' \ne k } 
 \frac{\epsilon_{k'}^2}{\epsilon^2_{k} - \epsilon^2_{k'}} \left(1 + G \sum_{k'' \ne k' }\frac{\epsilon^2_{k''}}{\epsilon^2_{k'} - \epsilon^2_{k''}} \right) \nonumber\\ && + \sum_{k' \ne k} \left[ \frac{G^2}{2}\left( \frac{\epsilon_{k}\epsilon_{k'}}{\epsilon^2_{k} - \epsilon^2_{k'}}\right)^2 
 +\frac{G^2}{4} \left( \frac{\epsilon_{k'}^2}{\epsilon^2_{k} - \epsilon^2_{k'}}\right)^2
 \right]
 \nonumber\\
 \nonumber\\
 && \tilde{R}^2_k - \tilde{R}_k   -  G \sum_{k' \ne k } \epsilon_{k'}^2
 \frac{\tilde{R}_k - \tilde{R}_{k'} }{\epsilon^2_{k} - \epsilon^2_{k'}} 
 - \left(\frac{\gamma}{\epsilon_k}\right)^2 
 =
-\frac{G^2}{4} \sum_{k' \ne k } \sum_{k'' \ne k }\frac{\epsilon^2_{k'}}{\epsilon^2_{k} - \epsilon^2_{k'}} \frac{\epsilon^2_{k''}}{\epsilon^2_{k} - \epsilon^2_{k''}} 
\nonumber\\ && \ +  \frac{G^2}{2} \sum_{k' \ne k } \sum_{k'' \ne k' }
 \frac{\epsilon_{k'}^2}{\epsilon^2_{k} - \epsilon^2_{k'}}  \frac{\epsilon^2_{k''}}{\epsilon^2_{k'} - \epsilon^2_{k''}}  + \sum_{k' \ne k} \left[ \frac{G^2}{2}\left( \frac{\epsilon_{k}\epsilon_{k'}}{\epsilon^2_{k} - \epsilon^2_{k'}}\right)^2 
 +\frac{G^2}{4} \left( \frac{\epsilon_{k'}^2}{\epsilon^2_{k} - \epsilon^2_{k'}}\right)^2
 \right]
  \nonumber\\
  \label{tempo}
\eea

On the right hand side the terms $k'' = k'$ of the first double sum and $k''=k$ in the second one respectively cancel the fourth and third (single sums) terms, so that it reduces to:

\bea
&&-\frac{G^2}{4} \sum_{k' \ne k } \sum_{k'' \ne k,k' } \left[  \frac{\epsilon^2_{k'}}{\epsilon^2_{k} - \epsilon^2_{k'}} \frac{\epsilon^2_{k''}}{\epsilon^2_{k} - \epsilon^2_{k''}} -2
 \frac{\epsilon_{k'}^2}{\epsilon^2_{k} - \epsilon^2_{k'}}  \frac{\epsilon^2_{k''}}{\epsilon^2_{k'} - \epsilon^2_{k''}}
\right]
\nonumber\\
&& \ =
-\frac{G^2}{4} \sum_{k' \ne k } \sum_{k'' \ne k,k' } \epsilon^2_{k'} \epsilon^2_{k''} \left[ \left( \frac{1}{\epsilon^2_{k} - \epsilon^2_{k'}} - \frac{1}{\epsilon^2_{k} - \epsilon^2_{k''}} \right) \frac{1}{\epsilon^2_{k'}-\epsilon^2_{k''}}-2 
 \frac{1}{\epsilon^2_{k} - \epsilon^2_{k'}}  \frac{1}{\epsilon^2_{k'} - \epsilon^2_{k''}}
\right]
\nonumber\\
&& \ =
-\frac{G^2}{4} \sum_{k' \ne k } \sum_{k'' \ne k,k' } \epsilon^2_{k'} \epsilon^2_{k''} \left[  - \frac{1}{\epsilon^2_{k} - \epsilon^2_{k''}} \frac{1}{\epsilon^2_{k'}-\epsilon^2_{k''}}- 
 \frac{1}{\epsilon^2_{k} - \epsilon^2_{k'}}  \frac{1}{\epsilon^2_{k'} - \epsilon^2_{k''}}
\right]
\nonumber\\
&& \ =
-\frac{G^2}{4} \sum_{k' \ne k } \sum_{k'' \ne k,k' } \epsilon^2_{k'} \epsilon^2_{k''} \left[  - \frac{1}{\epsilon^2_{k} - \epsilon^2_{k'}} \frac{1}{\epsilon^2_{k''}-\epsilon^2_{k'}}- 
 \frac{1}{\epsilon^2_{k} - \epsilon^2_{k'}}  \frac{1}{\epsilon^2_{k'} - \epsilon^2_{k''}}
\right] =0
\eea

Consequently eq. (\ref{tempo}) reduces to:

\bea
&& \tilde{R}^2_k - \tilde{R}_k   =  G \sum_{k' \ne k } \epsilon_{k'}^2
 \frac{\tilde{R}_k - \tilde{R}_{k'} }{\epsilon^2_{k} - \epsilon^2_{k'}} 
 + \left(\frac{\gamma}{\epsilon_k}\right)^2,
\eea

\noindent confirming Claeys  et al.'s result given in eq. (\ref{quadclaeys}).

\section{Conclusion}

In this work we have demonstrated how one can completely circumvent the Bethe ansatz procedure to determine the spectrum of integrable models of the spin-1/2 RG family. Independently of the symmetries of the chosen models, be they of the XXX, XXZ or XYZ-type, be they skew-symmetric or not, the integrability conditions, which the couplings need to satisfy in order to define the set of conserved charges, are sufficient to insure the existence of a closed set of quadratic relations between the commuting operators. Their eigenvalue spectrum can therefore be found as the various solutions of a ensemble of quadratic equations linking them. 

The approach requires nothing more than integrability, here defined as the existence of commuting conserved charges, and therefore applies to any member of the class studied without any requirement for $U(1)$ symmetry, for the existence of a properly defined pseudo-vacuum state, or for a known and usable Bethe ansatz approach to its solvability. In this sense, the difficulties which make XYZ models more complicated than XXZ or XXX models are completely lifted since the proposed construction provides, in an identical fashion, a set of quadratic Bethe equations for the conserved charges' eigenvalues.

The specification of the eigenvalues does not, by itself, provide an explicit representation of the corresponding eigenstate. However, in some models where an explicit Bethe ansatz solution has also been built, determinant representation for the scalar product of the eigenstate with an arbitrary tensor product of $S^z_i$ eigenstates: $\prod_{j=1}^M S^+_{i_j} \ket{\downarrow \downarrow \dots \downarrow }$ have actually been constructed \cite{faribaultschurichtdet,claeysdet,claeysdet2,tschirhartdet3} and can ultimately provide an approach for explicitly constructing the eigenstates. Whether a generic, non-model-dependant, approach can allow the reconstruction of eigenstates exclusively from the knowledge of the corresponding eigenvalues remains, for now, an open question.

\end{document}